# Interlocking nodal chains and their examples in carbon networks

Zhiwei Li[1,2], Yuee Xie[1,2]*, Po-Yao Chang[3,4], and Yuanping Chen[1,2]*

[1]Faculty of Science, Jiangsu University, Zhenjiang, 212013, Jiangsu, China
[2]School of Physics and Optoelectronics, Xiangtan University, Xiangtan, 411105, Hunan, China
[3]Max Planck Institute for the Physics of Complex Systems, Nöhnitzer Strasse 38, D-01187 Dresden, Germany
[4]Department of Physics, National Tsing Hua University, Hsinchu 30013, Taiwan

## Abstract

Nodal chain is a typical topological phase in nodal line semimetals. Here, we propose a new topological phase --- interlocking nodal chains, in which two sets of nodal chains are interlocked each other. It includes one- (1D), two- (2D) and three-dimensional (3D) versions, which can be produced by a three-band model. The 2D and 3D interlocking nodal chains will evolve into some other phases, such as double concentric isolated (or intersecting) nodal rings, coexisting nodal chain and isolated (or intersecting) nodal rings. These phases exhibit diverse surface states and Landau levels, which implies that there are rich electronic and magnetic properties associating with them. Moreover, the 2D interlocking nodal chains and related phase transitions can be realized in a series of carbon networks under strain. Without strain, topological phase in the carbon structures is double concentric isolated nodal rings. A larger tensile strain leads to the phase transiting to an interlocking nodal chain, while the middle phase is a coexisting phase of a nodal chain and isolated nodal rings. In addition, stability and synthesis of the carbon networks are discussed.

*Corresponding authors: xieyech@xtu.edu.cn; chenyp@ujs.edu.cn.

# 1. Introduction

Topological semimetal/metal (TM) has been a research focus in last several years [1-3] . On contrast to topological insulator, where no energy bands exist around the Fermi level, conduction and valence bands of TMs cross the Fermi level [4]. According to the cases of the band crossings, TMs have various classifications [5-22]. For example, if the crossings form points, lines or surfaces in the momentum space, the corresponding topological phases are dubbed as nodal points [5-7], nodal lines [8-19] or nodal surfaces [20, 21], respectively. Every topological phase has its own unique low-energy excitation and topological characteristics such as surface states [23-25] and magnetic properties [26-29]. Therefore, researchers are trying hard on one hand to search new topological phases, on the other hand to reveal novel topological properties and promising potentials for future applications.

Compared with nodal points and nodal surfaces, topological phases consisting of nodal lines have more subcategories because a line can be deformed into many different geometries [30-33] (e.g. a ring or a knot). If there are two or more than two lines/rings in the momentum space, they will lead to rich topological phases, such as nodal chains [34-38], nodal links [39, 40], Hopf chains [41-45], etc. In this sense, the topological elements can be viewed as building blocks, and new topological phases can be constructed based on these blocks. Nodal chains, where (orthogonal) nodal rings contact one by one periodically and then form chains along one or more directions, are typical phases in nodal line TMs [34, 35]. Figure 1(a) exhibits a nodal chain extending along one direction in the three-dimensional (3D) momentum space. It seems like a one-dimensional (1D) chain in the 3D space, and thus is named 1D nodal chain. More nodal chains can form two-dimensional (2D) and 3D networks by interconnections between them. Figure 1(b) exhibits an example of 2D nodal chains, where the chains extend along two directions and overspread on a 2D plane. Figure 1(c) is a 3D nodal chain where the chains extend along three mutually vertical directions. The contact points in the chains must be protected by special symmetries [16, 34]. Therefore, topological characteristics of nodal chains are different from topological characteristics of the non-contact nodal rings [16, 24, 34]. Experimentally, a nodal chain has been observed in a metallic-mesh photonic crystal [46].

A nodal chain is constructed by nodal rings. Mathematically, one can construct complicated graphs based on nodal chains. Figure 1(d) presents a 1D case where the red and blue nodal chains are interlocked each other. We refer this phase to interlocking nodal chains. Figures 1(e-f) show 2D and 3D versions of interlocked nodal chains corresponding to the nodal chains in Figs. 1(b-c), respectively. These graphs not only consist of nodal chains but also contain nodal links or Hopf links. However, can these interlocked nodal chains be new topological phases existing in momentum space? If yes, can one find these new phases in real materials?

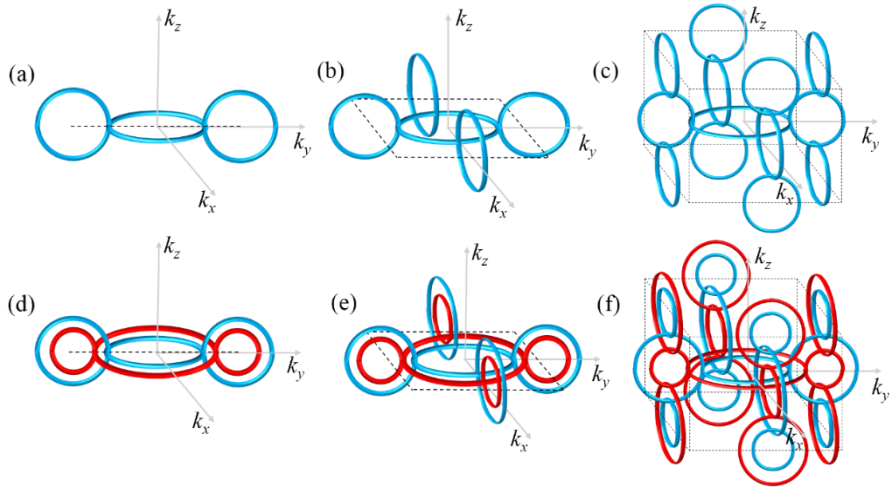

Figure 1. (a) 1D, (b) 2D and (c) 3D nodal chains. (d) 1D, (e) 2D and (f) 3D interlocking nodal chains where two sets of (blue and red) nodal chains are interlocked each other. The dashed lines, rectangles and cuboids represent periodic units of the topological phases.

Here, we propose that 1D, 2D and 3D interlocking nodal chains can be found in momentum space, which can be described by a three-band tight-binding model. The tight-binding model indicates that the 2D and 3D interlocking nodal chains can transit to various topological phases, and rich topological surface states and diverse Landau level spectra are found in these phases. Moreover, these topological phases and phase transitions can be realized in a series of carbon networks. Strain applied on the structures will lead to phase transitions, from double concentric nodal rings to coexisting of a nodal chain and isolated nodal rings, and then to interlocking nodal chains. We discuss the evolutions of topological surface states and synthesis of the carbon networks.

## 2. Tight-binding models for interlocking nodal chains and related phases

It is known that a 1D nodal chain in Fig. 1(a) can be described by a 2×2 tight-binding model as follow [44]:

$$H = \begin{pmatrix} H_{11} & H_{12} \\ H_{21} & H_{22} \end{pmatrix},$$
(1)

with

$$H_{11} = \Delta_1 + (A_1 cos(kx) + B_1 cos(ky) + C_1 cos(kz))/2,$$

$$H_{22} = \Delta_2 + (A_2 cos(kx) + B_2 cos(ky) + C_2 cos(kz))/2,$$
(2)

$$H_{12} = H_{21}^* = iDsin(ky/2)sin(kz/2),$$

where $\Delta_1, \Delta_2, A_1, A_2, B_1, B_2, C_1, C_2$ and $D$ are parameters. We find that Eq. (1) can also be used to describe the 2D nodal chain in Fig. 1(b) and 3D nodal chain in Fig. 1(c), by changing the off-diagonal terms to $H_{12} = H_{21}^* = iDsin(kx/2)sin(ky/2)sin(kz/2)$ and $iDsin(kx)sin(ky)sin(kz)$, respectively.

Equation (1) generates not only nodal chains but also some other related topological phases, by tuning its parameters. Figures 2(a-c) present the band structures produced by Eq. (1) with different parameters, and the corresponding phases are given in Figs. 2(d-f), respectively. In Fig. 2(a), two bands only cross along Γ-S and S-M. Its corresponding topological phase is isolated nodal rings located on the plane $k_z = 0$, as shown in Fig. 2(d). The two crossing points in Fig. 2(a) belong to the nodal ring with a center at point S. When the parameter $\Delta_1/\Delta_2$ decreases/increases, the black band shifts up while the blue band shifts down. This leads to the nodal rings in Fig. 2(d) becoming larger gradually. After the black band moves to the above of the blue band at point M [see Fig. 2(b)], phase transition occurs. The isolated nodal rings evolve into a 2D nodal chain in Fig. 2(e), where one ring locates on the plane $k_z = 0$ with a center at point Γ and other two vertical rings locate on the planes $k_x = 0$ and $k_y = 0$, respectively. When the two bands continue shifting, the ring on the plane $k_z = 0$ becomes smaller while those on the planes $k_{x/y} = 0$ become larger. If the black band shifts to the above of the blue band at the points Γ and T [see Fig. 2(c)], all the original crossing points of two bands disappear, while new crossing points appear along R-U and U-T. This generates a topological phase in Fig. 2(f). It is a phase of intersecting nodal rings, where two nodal rings locating on the planes $k_x = 0$ and $k_y = 0$ intersecting each other and share a center point U. Seen from Figs. 2(d-f), a nodal chain can transit to isolated nodal rings or intersecting nodal rings.

It is noted that all the nodal rings mentioned above are protected by mirror/glide planes. According to Eq. (1), the system has three mutually perpendicular mirror planes $M_x$, $M_y$ and $M_z$ satisfying $[M_i, H] = 0$, ($i$=x, y, z). The three mirror planes locate on the planes $k_x = 0$, $k_y = 0$ and $k_z = 0$, respectively, and each mirror operator has two eigenvalues of ±1 because $M_i{}^2 = 1$. In Figs. 2(a-c), the eigenvalues ±1 of mirror planes $M_x$, $M_y$ and $M_z$ are labeled as X±, Y± and Z±, respectively. One can find that the eigenvalues for the blue and black energy bands are always opposite, and thus the two bands can cross together. In Fig. 2(a), the band crossings form isolated rings on the plane $k_z = 0$ [see Fig. 2(d)], which are protected by mirror plane $M_z$. The nodal rings on the planes $k_x = 0$, $k_y = 0$ and $k_z = 0$ in Fig. 2(e) are protected by mirror planes $M_x$, $M_y$ and $M_z$, respectively. Moreover, the mirror planes are normal each other, and thus their operators are commute each other, i.e., $[M_x, M_z] = 0$ and $[M_y, M_z] = 0$. In this case, 2D nodal chains can form along $k_y$ and $k_z$ axes [16]. For the same reason, the intersecting nodal rings in Fig. 2(f) can form. The topological protection of all the nodal rings in Figs. 2(d-f) can also be inferred from their 1D winding numbers along a close path $\mathcal{L}$ encircling the rings: $N_\mathcal{L} = \frac{1}{\pi} \oint_\mathcal{L} d\mathbf{k} \cdot A(\mathbf{k})$, where $A(\mathbf{k})$ is the Berry connection at the point $\mathbf{k}$. The calculation results indicate that all of them have nontrivial values.

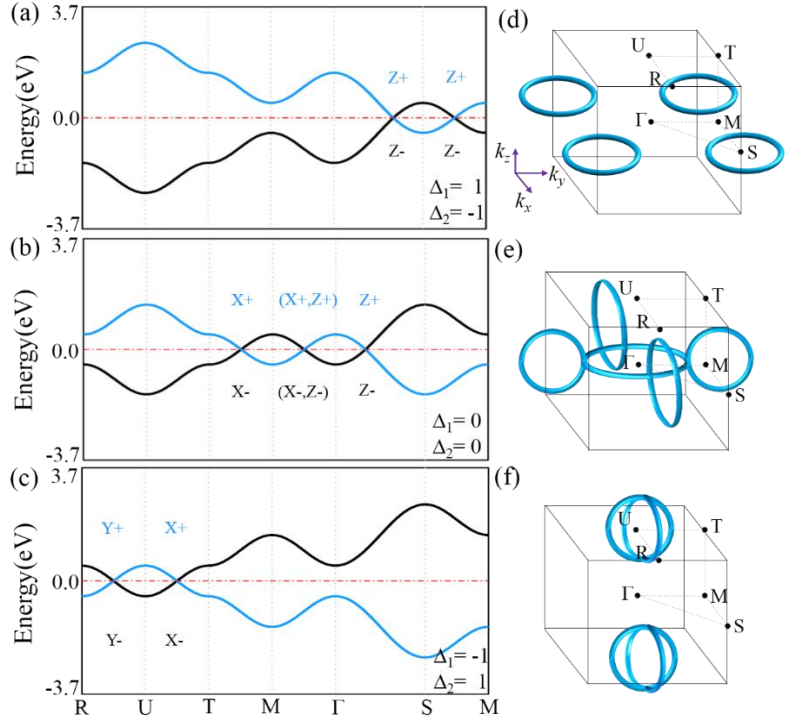

Figure 2. (a-c) Band structures based on Eq. (1) for different parameters $\Delta_1$ and $\Delta_2$ ($A_1$ = 1, $B_1$ = 1.02, $C_2$ = -1.1, $A_2$ = -1, $B_2$ = -1.02, $C_1$ = 1.1, $D$ = 1). (a) $\Delta_1$ = 1, $\Delta_2$ = -1; (b) $\Delta_1$

$= \Delta_2 = 0$; (c) $\Delta_1 = -1, \Delta_2 = 1$. Topological phases from (d) isolated nodal rings to (e) a nodal chain to (f) intersecting nodal rings, which correspond to the band structures in (a-c), respectively. The eigenvalues of mirror planes $k_x = 0$, $k_y = 0$ and $k_z = 0$ are labeled as X±, Y± and Z± in (a-c).

One can extend Eq. (1) to a three-band model:

$$H = \begin{pmatrix} H_{11} & H_{12} & 0 \\ H_{21} & H_{22} & H_{23} \\ 0 & H_{32} & H_{33} \end{pmatrix}, \quad (3)$$

with

$$H_{11} = \Delta_1 + (A_1 cos(kx) + B_1 cos(ky) + C_1 cos(kz))/2,$$

$$H_{22} = \Delta_2 + (A_2 cos(kx) + B_2 cos(ky) + C_2 cos(kz))/2,$$

$$H_{33} = \Delta_3 + (A_3 cos(kx) + B_3 cos(ky) + C_3 cos(kz))/2, \quad (4)$$

$$H_{12} = H_{21}^* = iD_1 sin(ky/2)sin(kz/2),$$

$$H_{23} = H_{32}^* = iD_2 sin(ky/2)sin(kz/2),$$

where $\Delta_1$, $\Delta_2$, $\Delta_3$, $A_1$, $A_2$, $A_3$, $B_1$, $B_2$, $B_3$, $C_1$, $C_2$, $C_3$, and $D_1$, $D_2$ are parameters. It can generate a 1D interlocking nodal chain in Fig. 1(d). If one changes the off-diagonal terms to

$$H_{12} = H_{21}^* = iD_1 sin(kx/2)sin(ky/2)sin(kz/2), \quad (5)$$

$$H_{23} = H_{32}^* = iD_2 sin(kx/2)sin(ky/2)sin(kz/2),$$

Eq. (3) can generate a 2D interlocking nodal chain in Fig. 1(e). When $sin(kx/2)sin(ky/2)sin(kz/2)$ in Eq. (5) is further replaced by $sin(kx)sin(ky)sin(kz)$, Eq. (3) can generate a 3D interlocking nodal chain in Fig. 1(f) Similar to the cases of Eq. (1), Eq. (3) can produce topological phases other than interlocked nodal chains by changing the parameters. Figures 3(a-e) shows five typical phases produced by Eq. (3) with different $\Delta_2$, and their corresponding band structures are given in Fig. 3(f). Figure 3(a) is a phase of double concentric nodal rings (one is a blue ring and the other is a red ring), which is a double version of the phase in Fig. 2(d). Seen from Fig. 3(f), the blue/red ring is induced by the crossings of blue/red and black bands. When the black band shifts up, the blue nodal rings evolves into a nodal chain, which is like the phase transition from Figs. 2(a) to 2(b). It leads to a coexisting phase of a nodal chain and isolated nodal ring, as shown in Fig. 3(b). When the black band shifts up continuously, the red nodal rings also transit to a nodal chain, and then interlocking nodal chains are formed [see Fig. 3(c)]. Figure 3(d) is another coexisting phase of a nodal chain and intersecting nodal rings, which is evolved from Fig. 3(c) when the blue nodal chain changes to intersecting nodal rings. After the red nodal chain in Fig. 3(d) also transits to intersecting nodal rings, double concentric intersecting nodal rings are obtained as shown in Fig. 3(e).

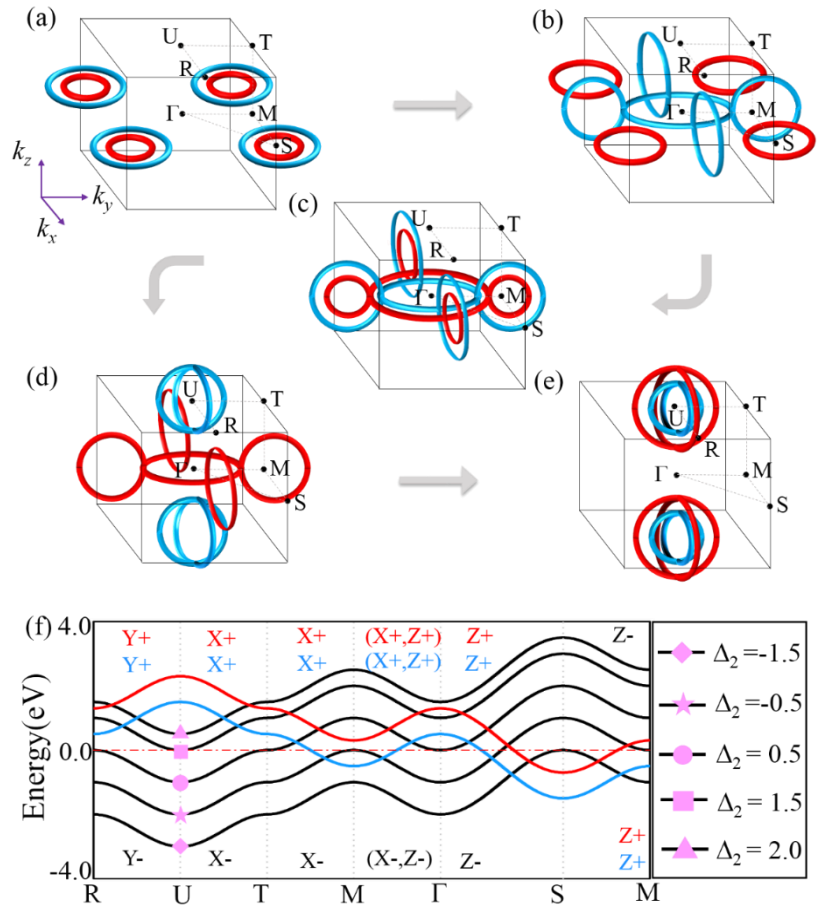

Figure 3. (a-e) Topological phases based on Eq. (3) for different parameter $\Delta_2$ ($A_1 = A_3$ =1, $B_1 = B_3$ =1.02, $C_1 = C_3$ = -1, $A_2 = -1$, $B_2 = -1.02$, $C_2 = 1$, $D_1 = D_2 = 1$, $\Delta_1 = 0$, $\Delta_3 =$ 0.8). (a) Double concentric nodal rings with $\Delta_2 = -1.5$; (b) coexisting of a nodal chain and nodal rings with $\Delta_2 = -0.5$; (c) interlocking nodal chains with $\Delta_2 = 0.5$; (d) coexisting of a nodal chain and intersecting nodal rings with $\Delta_2 = 1.5$; (e) double concentric intersecting nodal rings with $\Delta_2 = 2.0$. (f) Band structures corresponding to the five topological phases in (a-e), where X$\pm$, Y$\pm$ and Z$\pm$ label the eigenvalues of mirror planes $k_x = 0$, $k_y = 0$ and $k_z = 0$, respectively. It is noted that the variation of $\Delta_2$ leads to the shift of the black band in (f).

For the system described by Eq. (3), it also has three mutually perpendicular mirror planes $M_x$, $M_y$ and $M_z$ satisfying [$M_i$, $H$] =0, (i=x, y, z). Their eigenvalues corresponding to the three energy bands are given in Fig. 3(f). One can find that the red and blue bands have the same eigenvalues while the black band has opposite values. Therefore, the red and blued bands cannot cross each other. The crossings between red and black bands and those between blue and black bands are independent. Thus, one can use the same analysis of symmetry like that in Fig. 2 to explain topological protections of the phases in Figs. 3(a-e).

## 3. Topological surface states and Landau levels

Interlocking nodal chains and related topological phases possess diverse topological surface states. Figure 4 presents surfaces states in the first BZ for the topological phases from Figs. 3(a) to 3(e). The surface states on surface [001] are given in the top panel, while those on surface [100] are given in the bottom panel. Each phase has its unique surface state.

On the surface [001], the surface state for the double concentric nodal rings appears at the corner of the first BZ, moreover, there are two surface states (yellow region) existing in the inner projected ring [see Fig. 4(a)]. When the phase transits to a nodal chain and isolated nodal rings, the (pink) region containing one surface state expands toward the center, as shown in Fig. 4(b). When the phase further transits to interlocking nodal chains, the yellow region with two surface states also expand toward the center [see Fig. 4(c)]. In some case, two surface states can nearly overspread the whole BZ. After the interlocking nodal chains evolve into a nodal chain and intersecting nodal rings, the yellow region disappears while the pink region nearly paves the BZ except a center ring [see Fig. 4(d)]. It means that one surface state nearly occupies the whole BZ. Finally, all surface states will disappear if the phase evolves into double concentric intersecting nodal rings [see Fig. 4(e)]. Seen from Figs. 4(a) to 4(e), one can find not only diverse surface states but also diverse evolution processes between them. The evolution of surface states on [100] surface exhibits richer configurations. These results indicate that interlocking nodal chains and related phase transitions possess exotic electronic transport properties different from those in conventional Weyl/Dirac semimetals.

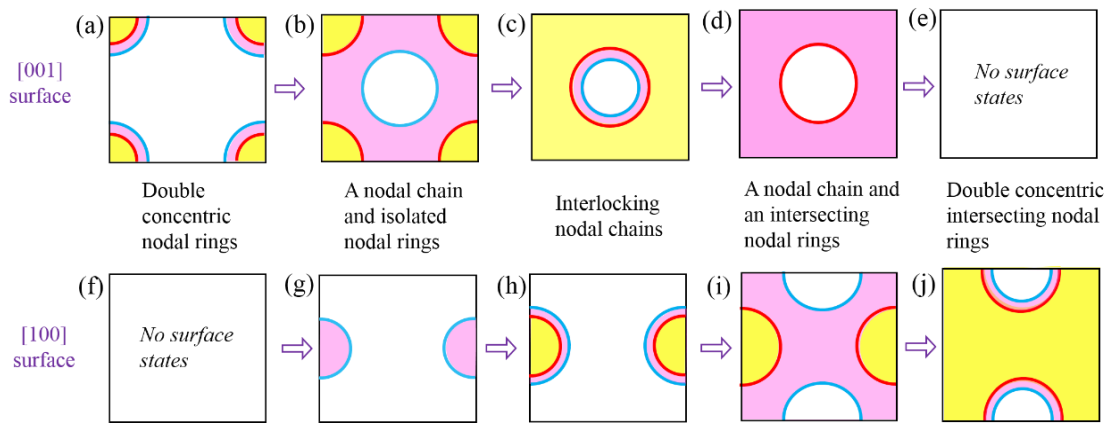

Figure 4. Topological surface states corresponding to the topological phases in Figs.

3(a)-(e). (a)-(e) The evolution of surface states on the [001] surface in the first BZ. (f)-(j) The evolution of surface states on the [100] surface in the first BZ. The white, pink and yellow regions represent none, one and two surface states, respectively.

Topological surface states reveal 2D topological characteristics of topological phases, while 1D characteristics can be revealed by Landau level. We expand the effective tight-binding model at $\Gamma$ point and write the conjugate momenta $\Pi_\alpha = k_\alpha - eA_\alpha = \frac{1}{\sqrt{2}l_B}(a^\dagger + a)$ and $\Pi_\beta = k_\beta - eA_\beta = \frac{1}{\sqrt{2}il_B}(a^\dagger - a)$, but keep $\gamma$ direction periodic, i.e., in Eq. (3), $sink_\gamma$ and $cosk_\gamma$ are unchanged. Here $l_B = 1/\sqrt{eB}$ is the magnetic length with $B$ is the magnetic field along $\gamma$ direction, $(\alpha, \beta, \gamma)$ form a cyclic permutation of $(x, y, z)$, and $(a, a^\dagger)$ are the ladder operators. The Landau level spectrum is solved numerically from the eigen-equations $H(k_\gamma, B)\Psi = E(k_\gamma, B)\Psi$, where $\Psi = (\sum_n \alpha_n |n\rangle, \sum_n \beta_n |n\rangle, \sum_n \gamma_n |n\rangle)$ with $a^\dagger |n\rangle = \sqrt{n+1}|n+1\rangle$ and $a|n\rangle = \sqrt{n}|n-1\rangle$. The Landau level spectra for the interlocking nodal chains and related phases are shown in Fig. 5. One should be noticed that since we only expand the tight-binding model at $\Gamma$ point along $\gamma$ direction, the nodal lines far away from the $\Gamma$ point along $\gamma$ direction cannot be probed by the Landau level spectrum. It is shown in a generic two-band model [47], the nodal ring TMs can exhibit non-dispersive (flat) Landau levels as a function of the momentum along the magnetic field direction, depending on the nodal ring projection on the magnetic field direction. However, in a three-band model, the Landau level spectrum is rather complicated. The reason of the complication comes from the level repulsion. The Landau levels spread out from different bulk bands can be either upward or downward along the energy axis. In the situation that the Landau levels of the upper band going downward and the Landau levels of the lower band going upward, the level repulsion happens. This level repulsion can flatten the dispersion for large Landau level index [see the light blue levels between two bulk bands in Fig. 5]. These flat bands with large Landau level indices are most located within regions that the nodal rings projected onto. When the magnetic field is along z-direction, one can find flat bands in the region where the nodal ring projected onto [Figs. 5(d-e)]. Similarly, when the magnetic field is along x-direction, there are flat bands in the region where the nodal rings projected onto [Figs. 5(g-i)]. For the zeroth Landau levels, they follow the original bulk spectrum but will be gapped at the crossings. This is very different from the crossing in the Weyl TMs, where the zeroth Landau level will remain gapless at the crossings. The existence of a large number of flat bands (Landau levels with large Landau level indices) near the Fermi level can has significant contribution to the density of states which can be probed by scanning tunneling microscopy. Since the existence of these flat bands depends on the projection of the nodal-ring striation on the magnetic field direction, it provides an indirect way to probe the structure of the interlocking nodal chains.

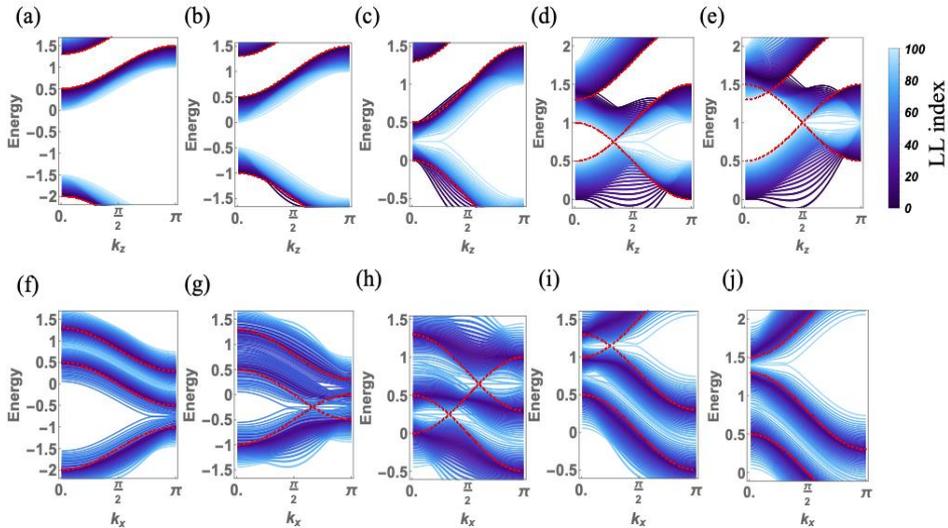

Figure 5. Landau level spectrum as a function of $k_z$ (magnetic field along z-direction) (a-e) and as a function of $k_x$ (magnetic field along $x$-direction) (f-j), with the magnetic

field $eB = 0.01$. The parameters in the tight-binding model are the same as in Fig. 3: (a) and (f) correspond to Fig. 3(a); (b) and (g) correspond to Fig. 3(b); (c) and (h) correspond to Fig. 3(c); (d) and (i) correspond to Fig. 3(d); (e) and (j) correspond to Fig. 3(e). The color scale is the Landau level (LL) index. The dashed red lines are the bulk spectrum with zero magnetic field.

## 4. Realization of interlocking nodal chains in carbon networks

The interlocking nodal chains and related topological phases can be realized in a series of carbon networks, consisting of armchair graphene nanoribbons, as shown in Fig. 6(a). The atoms in the nanoribbons are divided into two groups, purple C1 atoms at the edges and inner light-blue C2 atoms. $n$ is used to label width of the region with C2 atoms. Figure 6(b) presents atomic structure of a carbon network, where the nanoribbons are connected according to triangle lattice pattern, in another words, each nanoribbon is equal to a site in the triangle lattice. In the 3D network, the C1 atoms become $sp^3$ hybridized, while C2 atoms are still $sp^2$ hybridized. Because top view of the 3D network looks like a γ-graphyne, we name it γg-AGN-$n$. The variation of $n$ produces a series of structures. The structure in Fig. 6(b) is γg-AGN-2, whose primitive cell is shown in Fig. 6(d). Figure 6(c) presents the first BZ of the primitive cell.

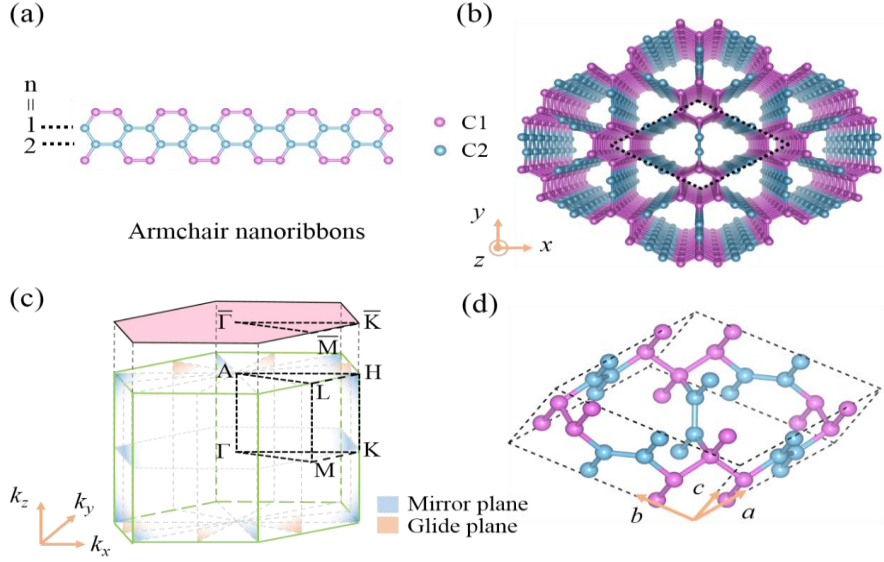

Figure 6. (a) Armchair nanoribbons: *n* defines the ribbon widths of light blue atom. (b) Top perspective view of γg-AGN-2 which consists of nanoribbons in (a). The light-blue C2 atoms are $sp^2$ hybridized, while the pink C1 atoms are $sp^3$ hybridized. (c) The first BZ and its projection on the [001] surface, where four mirror planes and three glide planes are shown. (d) Side view of the primitive cell of γg-AGN-2, and its top view is given in the dashed box in (b). *a*, *b* and *c* represent lattice constants.

We performed first-principles calculations within the DFT formalism as implemented in VASP code [48]. The potential of the core electrons and the exchange-correlation interaction between the valence electrons are described, respectively, by the projector augmented wave approach [49] and the generalized gradient approximation (GGA) with Perdew–Burke–Ernzerhof (PBE) functional [50]. The kinetic energy cutoff of 500 eV is employed. The atomic positions were optimized using the conjugate gradient method. The energy and force convergence criteria are set to be $10^{-6}$ eV and $10^{-3}$ eV/Å, respectively. The phonon calculation is carried out using the Phonopy package [51]. The band structures are calculated using the primitive cell. The Brillouin zone was sampled with a $7 \times 7 \times 11$ Monkhorst–Pack special k-point grid for geometrical optimization. The surface states calculations were calculated by the open-source software Wannier-Tools package [52].

The space group of γg-AGN-2 is P6$_3$/mcm, including one mirror plane normal to $z$ axis, three mirror planes and three glide planes parallel to $z$ axis [see Fig. 6(c)]. The mirror/glide planes satisfy C$_3$ rotation symmetry. The density of γg-AGN-2 is between diamond and graphite. The bond lengths are in the range of 1.37 ~ 1.58 Å, and most of them are between those of diamond (1.54 Å) and graphite (1.42 Å). The calculated cohesive energy E$_{coh}$ shows that γg-AGN-2 (-7.59 eV/atom) is a metastable allotrope, only 0.18 eV/atom smaller than diamond but greater than some other carbon allotropes such as CKL [53] and CHC-2 [54]. [Detail structure parameters can be seen in Table S1 in Supplementary Information (SI)]. To clarify stability of γg-AGN-2, we calculate its phonon dispersion, as shown in Fig. S1. No imaginary frequency is found, which indicates that it has a good stability.

The band structure for γg-AGN-2 is calculated, and the result is shown in Fig. 7(a). One can find that there are three bands around the Fermi level, and they only cross along K-M and K-Γ. The crossing points lie on two nodal rings on the plane $k_z = 0$, respectively, as shown in Fig. 7(d). This is a phase of double concentric nodal rings, similar to that in Fig. 3(a). When a strain is applied on $z$ axis, the red and blue bands shift down while the black band shift up [see Fig. 7(b)]. Similar to the case in Fig. 3, the double concentric nodal rings evolve into a coexisting phase of a nodal chain and isolated nodal ring, as shown in Fig. 7(e). When a larger strain is applied on the structure, the bands continue shifting [see Fig. 7(c)]. Correspondingly, the phase transits to interlocking nodal chains, as shown in Fig. 7(f). All the nodal rings in Figs. 7(d-f) are protected by the mirror plane on $k_z = 0$ and three glide planes. The phase transitions from Figs. 7(d) to 7(f) are similar to those from Figs. 3(a) to 3(c). The difference between them is the shapes of the BZs. Therefore, the evolution of [001] surface states for γg-AGN-2 is also similar to Figs. 4(a-c) [see Fig. S2 in SI].

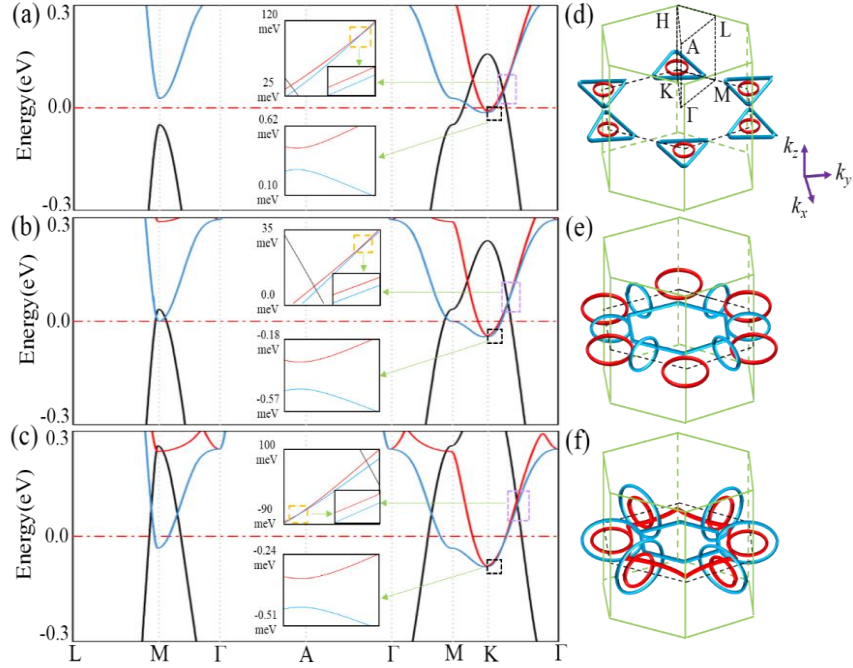

Figure 7. Band structures for γg-AGN-2 in the cases of (a) no strain, (b) 1% tensile strain and (c) 3% tensile strain along $z$ axis. Insets: amplified regions of the purple, black and orange dashed boxes, respectively. (d-f) Three topological phases corresponding to band structures in (a-c), respectively. (d) is a phase of double concentric nodal rings, (e) is a coexisting phase of a nodal chain and isolated nodal rings, and (f) is interlocking nodal chains. The difference between the phases in (d-f) and those in Figs. 3(a-c) is the former distribute in a hexagonal BZ while the latter distribute in a rectangular BZ.

Beside γg-AGN-2, the family of structures γg-AGN-6$m$+2 ($m$ is an integer) can share these topological phases and phase transitions mentioned above. This is because all the structures have the same crystal symmetry and the graphene nanoribbons with width 6$m$+2 have similar electronic properties. Figure S3 in SI shows the structure of γg-AGN-8 and their band structures under strain. One can find the evolution of band structures under strain is very similar to the case in γg-AGN-2.

The series of structures γg-AGN-$n$ could be synthesized by squeezing a bundle of carbon nanotubes. The left panel in Fig. S4 in SI presents atomic structure of the nanotube bundle, which is a stable close-packed structure. Its lattice constants are $a_0 = b_0 = 7.72$ Å. A calculation of structural optimization indicates that a hydrostatic pressure approximates 25 GPa will lead to the bundle deforming into a compressive γg -AGN whose lattice constants are $a_1 = b_1 = 6.71$ Å. When the pressure is released, the compressive structure relaxes to a regular γg-AGN. Figure S4 in SI shows the process of structural transition from nanotube bundle to carbon network. The energy barrier needs to overcome in this process is 0.051eV/atom.

## 5. Conclusions

In conclusion, we propose a new type of topological phases --- interlocking nodal chains, where two (or more) sets of nodal chains are interlocked together. A three-band tight-binding model reveals that the new phases can be evolved from some other topological phases, such as double concentric isolated/intersecting nodal rings, coexisting phases of a nodal chain and isolated/intersecting nodal rings. These phases exhibit diverse topological surface states and Landau levels. We also find that the phases and phase transitions can be realized in a series of carbon networks, which could be synthesized by squeezing a bundle of carbon nanotubes.

## Acknowledgments

This work was supported by the National Natural Science Foundation of China (No. 11874314).

# References

[1] A.A. Burkov, Topological semimetals, Nat. Mater. 15 (2016) 1145.

[2] W. Hongming, D. Xi, F. Zhong, Topological semimetals predicted from first-principles calculations, J. Phys.: Condens. Matter 28(30) (2016) 303001.

[3] H. Gao, J.W.F. Venderbos, Y. Kim, A.M. Rappe, Topological Semimetals from First Principles, Ann. Rev. Mater. Res. 49(1) (2019) 153-183.

[4] S.-Y. Yang, H. Yang, E. Derunova, S.S.P. Parkin, B. Yan, M.N. Ali, Symmetry demanded topological nodal-line materials, Adv. Phys.: X 3(1) (2018) 1414631.

[5] B. Bradlyn, J. Cano, Z. Wang, M.G. Vergniory, C. Felser, R.J. Cava, et al., Beyond Dirac and Weyl fermions: Unconventional quasiparticles in conventional crystals, Science 353(6299) (2016).

[6] J. Hu, W. Wu, C. Zhong, N. Liu, C. Ouyang, H.Y. Yang, et al., Three-dimensional honeycomb carbon: Junction line distortion and novel emergent fermions, Carbon 141 (2019) 417-426.

[7] A.A. Soluyanov, D. Gresch, Z. Wang, Q. Wu, M. Troyer, X. Dai, et al., Type-II Weyl semimetals, Nature 527 (2015) 495.

[8] Y. Chen, Y. Xie, S.A. Yang, H. Pan, F. Zhang, M.L. Cohen, et al., Nanostructured Carbon Allotropes with Weyl-like Loops and Points, Nano Lett. 15(10) (2015) 6974-6978.

[9] T. Nomura, T. Habe, R. Sakamoto, M. Koshino, Three-dimensional graphdiyne as a topological nodal-line semimetal, Phys. Rev. Mater. 2(5) (2018) 054204.

[10] H. Zhang, Y. Xie, Z. Zhang, C. Zhong, Y. Li, Z. Chen, et al., Dirac Nodal Lines and Tilted Semi-Dirac Cones Coexisting in a Striped Boron Sheet, J. Phys. Chem. Lett. 8(8) (2017) 1707-1713.

[11] C. Niu, P.M. Buhl, G. Bihlmayer, D. Wortmann, Y. Dai, S. Blügel, et al., Two-dimensional topological nodal line semimetal in layered $X_2Y$ (X = Ca, Sr, and Ba; Y = As, Sb, and Bi), Phys. Rev. B 95(23) (2017) 235138.

[12] Y. Chen, Y. Xie, Y. Gao, P.-Y. Chang, S. Zhang, D. Vanderbilt, Nexus networks in carbon honeycombs, Phys. Rev. Mater. 2(4) (2018) 044205.

[13] Y. Cheng, J. Du, R. Melnik, Y. Kawazoe, B. Wen, Novel three dimensional topological nodal line semimetallic carbon, Carbon 98 (2016) 468-473.

[14] Y. Gao, Y. Chen, Y. Xie, P.-Y. Chang, M.L. Cohen, S. Zhang, A class of topological nodal rings and its realization in carbon networks, Phys. Rev. B 97(12) (2018) 121108.

[15] J.-T. Wang, H. Weng, S. Nie, Z. Fang, Y. Kawazoe, C. Chen, Body-Centered Orthorhombic $C_{16}$: A Novel Topological Node-Line Semimetal, Phys. Rev. Lett. 116(19) (2016) 195501.

[16] C. Gong, Y. Xie, Y. Chen, H.-S. Kim, D. Vanderbilt, Symmorphic Intersecting Nodal Rings in Semiconducting Layers, Phys. Rev. Lett. 120(10) (2018) 106403.

[17] R. Yu, H. Weng, Z. Fang, X. Dai, X. Hu, Topological Node-Line Semimetal and Dirac Semimetal State in Antiperovskite Cu3PdN, Phys. Rev. Lett. 115(3) (2015) 036807.

[18] Y. Gao, Y. Xie, Y. Chen, J. Gu, Z. Chen, Spindle nodal chain in three-dimensional α′ boron, Phys. Chem. Chem. Phys. 20(36) (2018) 23500-23506.

[19] D.-S. Ma, J. Zhou, B. Fu, Z.-M. Yu, C.-C. Liu, Y. Yao, Mirror protected multiple nodal line semimetals and material realization, Phys. Rev. B 98(20) (2018) 201104.

[20] C. Zhong, Y. Chen, Y. Xie, S.A. Yang, M.L. Cohen, S.B. Zhang, Towards three-dimensional Weyl-surface semimetals in graphene networks, Nanoscale 8(13) (2016) 7232-7239.

[21] J. Wang, Y. Liu, K.-H. Jin, X. Sui, L. Zhang, W. Duan, et al., Pseudo Dirac nodal sphere semimetal, Phys. Rev. B 98(20) (2018) 201112.

[22] Q.-F. Liang, J. Zhou, R. Yu, Z. Wang, H. Weng, Node-surface and node-line fermions from nonsymmorphic lattice symmetries, Phys. Rev. B 93(8) (2016) 085427.

[23] B.Q. Lv, H.M. Weng, B.B. Fu, X.P. Wang, H. Miao, J. Ma, et al., Experimental Discovery of Weyl Semimetal TaAs, Phys. Rev. X 5(3) (2015) 031013.

[24] Y. Zhou, F. Xiong, X. Wan, J. An, Hopf-link topological nodal-loop semimetals, Phys. Rev. B 97(15) (2018) 155140.

[25] H. Weng, Y. Liang, Q. Xu, R. Yu, Z. Fang, X. Dai, et al., Topological node-line semimetal in three-dimensional graphene networks, Phys. Rev. B 92(4) (2015) 045108.

[26] Z. Wang, M.G. Vergniory, S. Kushwaha, M. Hirschberger, E.V. Chulkov, A. Ernst, et al., Time-Reversal-Breaking Weyl Fermions in Magnetic Heusler Alloys, Phys. Rev. Lett. 117(23) (2016) 236401.

[27] E. Emmanouilidou, B. Shen, X. Deng, T.-R. Chang, A. Shi, G. Kotliar, et al., Magnetotransport properties of the single-crystalline nodal-line semimetal candidates CaT X (T=Ag, Cd; X=As, Ge), Phys. Rev. B 95(24) (2017) 245113.

[28] Z.-M. Yu, Y. Yao, S.A. Yang, Predicted Unusual Magnetoresponse in Type-II Weyl Semimetals, Phys. Rev. Lett. 117(7) (2016) 077202.

[29] H. Li, H. He, H.-Z. Lu, H. Zhang, H. Liu, R. Ma, et al., Negative magnetoresistance in Dirac semimetal $Cd_3As_2$, Nat. Commun. 7 (2016) 10301.

[30] L. Li, X. Kong, F.M. Peeters, New nanoporous graphyne monolayer as nodal line semimetal: Double Dirac points with an ultrahigh Fermi velocity, Carbon 141 (2019) 712-718.

[31] W. Chen, H.-Z. Lu, J.-M. Hou, Topological semimetals with a double-helix nodal link, Phys. Rev. B 96(4) (2017) 041102.

[32] R. Bi, Z. Yan, L. Lu, Z. Wang, Nodal-knot semimetals, Phys. Rev. B 96(20) (2017) 201305.

[33] Z. Yang, J. Hu, Non-Hermitian Hopf-link exceptional line semimetals, Phys. Rev. B 99(8) (2019) 081102.

[34] T. Bzdušek, Q. Wu, A. Rüegg, M. Sigrist, A.A. Soluyanov, Nodal-chain metals, Nature 538 (2016) 75.

[35] J. Cai, Y. Xie, P.-Y. Chang, H.-S. Kim, Y. Chen, Nodal-chain network, intersecting nodal rings and triple points coexisting in nonsymmorphic $Ba_3Si_4$, Phys. Chem. Chem. Phys. 20(32) (2018) 21177-21183.

[36] S.-S. Wang, Y. Liu, Z.-M. Yu, X.-L. Sheng, S.A. Yang, Hourglass Dirac chain metal in rhenium dioxide, Nat. Commun. 8(1) (2017) 1844.

[37] R. Yu, Q. Wu, Z. Fang, H. Weng, From Nodal Chain Semimetal to Weyl Semimetal in HfC, Phys. Rev. Lett. 119(3) (2017) 036401.

[38] Z. Yan, Z. Wang, Floquet multi-Weyl points in crossing-nodal-line semimetals, Phys. Rev. B 96(4) (2017) 041206.

[39] P.-Y. Chang, C.-H. Yee, Weyl-link semimetals, Phys. Rev. B 96(8) (2017) 081114.

[40] Z. Yan, R. Bi, H. Shen, L. Lu, S.-C. Zhang, Z. Wang, Nodal-link semimetals, Phys. Rev. B 96(4) (2017) 041103.

[41] C. Zhong, Y. Chen, Z.-M. Yu, Y. Xie, H. Wang, S.A. Yang, et al., Three-dimensional Pentagon Carbon with a genesis of emergent fermions, Nat. Commun. 8 (2017) 15641.

[42] Y. Xie, J. Cai, J. Kim, P.-Y. Chang, Y. Chen, Hopf-chain networks evolved from triple points, Phys. Rev. B 99(16) (2019) 165147.

[43] Z. Yang, C.-K. Chiu, C. Fang, J. Hu, Evolution of nodal lines and knot transitions in topological semimetals, arXiv:1905.00210.

[44] G. Chang, S.-Y. Xu, X. Zhou, S.-M. Huang, B. Singh, B. Wang, et al., Topological Hopf and Chain Link Semimetal States and Their Application to $Co_2MnGa$, Phys. Rev. Lett. 119(15) (2017) 156401.

[45] Q. Wu, A.A. Soluyanov, T. Bzdušek, Beyond the Tenfold Way: Non-Abelian Topology in Noninteracting Metals, arXiv:1808.07469.

[46] Q. Yan, R. Liu, Z. Yan, B. Liu, H. Chen, Z. Wang, et al., Experimental discovery of nodal chains, Nat. Phys. 14(5) (2018) 461-464.

[47] J.-W. Rhim, Y.B. Kim, Landau level quantization and almost flat modes in three-dimensional semimetals with nodal ring spectra, Phys. Rev. B 92(4) (2015) 045126.

[48] G. Kresse, J. Furthmüller, Efficiency of ab-initio total energy calculations for metals and semiconductors using a plane-wave basis set, Comp. Mater. Sci. 6(1) (1996) 15-50.

[49] G. Kresse, D. Joubert, From ultrasoft pseudopotentials to the projector augmented-wave method, Phys. Rev. B 59(3) (1999) 1758-1775.

[50] J.P. Perdew, K. Burke, M. Ernzerhof, Generalized Gradient Approximation Made Simple, Phys. Rev. Lett. 77(18) (1996) 3865-3868.

[51] A. Togo, F. Oba, I. Tanaka, First-principles calculations of the ferroelastic transition between rutile-type and $CaCl_2$-type $SiO_2$ at high pressures, Phys. Rev. B 78(13) (2008) 134106.

[52] Q. Wu, S. Zhang, H.-F. Song, M. Troyer, A.A. Soluyanov, WannierTools: An open-source software package for novel topological materials, Comput. Phys. Commun. 224 (2018) 405-416.

[53] Y. Chen, Y.Y. Sun, H. Wang, D. West, Y. Xie, J. Zhong, et al., Carbon Kagome Lattice and Orbital-Frustration-Induced Metal-Insulator Transition for Optoelectronics, Phys. Rev. Lett. 113(8) (2014) 085501.

[54] Y. Gao, Y. Chen, C. Zhong, Z. Zhang, Y. Xie, S. Zhang, Electron and phonon properties and gas storage in carbon honeycombs, Nanoscale 8(26) (2016) 12863-12868.